 \documentstyle[preprint,aps]{revtex}
\tighten
\include{epsf}
\newcommand{\bel}[1]{\begin{equation}\label{#1}}  
\newcommand{\bal}[1]{\begin{eqnarray}\label{#1}}
\newcommand{\be}{\begin{equation}}
\newcommand{\ee}{\end{equation}}
\newcommand{\ba}{\begin{eqnarray}}
\newcommand{\ea}{\end{eqnarray}}
\newcommand{\nn}{\nonumber\\}

\newcommand{\bes}{\begin{equation*}}
\newcommand{\ees}{\end{equation*}}
\newcommand{\del}{\partial}
\newcommand{\eqn}[1]{(\ref{#1})}
\newcommand{\mean}[1]{\langle #1 \rangle}
\newcommand{\hp}{\hat \phi}
\newcommand {\hk}{\widehat k}
\newcommand {\hq}{\widehat q}
\newcommand {\bp}{\overline \phi}
\newcommand {\wh}{\widehat}
\newcommand {\wt}{\widetilde}
\newcommand{\ov}{\overline}
\title{Optimization of Monte-Carlo calculations of the \\
\begin {centering}
 effective potential
\end {centering}
} 
\author{A. Ardekani$^{a,b\;\ast}$ and A. G. Williams$^{a,b\;\ddagger}$\\
$^a$Department of Physics and Mathematical Physics \\
$^b$ Special Research Center for the Subatomic Structure of Matter,\\
University of Adelaide, SA 5005, Australia}
\begin{document}
\preprint{\parbox[t]{50mm}{\flushright{Preprint: ADP-97-13/T250\\ 
   hep-ph/9705021}}}
\maketitle
\begin{abstract}
\normalsize
\noindent
We study Monte Carlo calculations of the
 effective potential for a scalar field theory using three techniques.
 One of these is a new method proposed and tested for the first time. 
In each case we  extract the
 renormalised quantities of the theory. The system studied in our
 calculations is a one component $\phi^4$ model in two dimensions. We apply 
these
 methods to  both the weak and strong coupling regime. In the weak coupling 
 regime  we compare
 our results for the renormalised quantities with those obtained
 from  two-loop lattice perturbation theory. Our results are verified in 
the strong coupling regime  through comparison with 
 the strong coupling expansion. We conclude that effective potential methods,
 when suitably chosen, can be accurate tools in calculations of the 
renormalised parameters of scalar field theories. 

\end{abstract}
\vspace{3cm}
\noindent\rule[0mm]{3cm}{0.2mm}
\vspace{1cm}\\
$^\ast$ {\tt aardekan@physics.adelaide.edu.au}\\
$^\ddagger$ {\tt awilliam@physics.adelaide.edu.au}\\
\newpage
\section {Introduction}
An understanding of the underlying vacuum structure of a quantum field 
theory is 
essential for  understanding its physical content. This analysis 
is conveniently carried out by calculating a quantity known as the 
effective potential~\cite{1,2,3},
denoted by $U(\overline\phi)$ and the 
minimum of which gives information as to the nature of the lowest energy 
eigenstate of the theory. This makes $U(\overline\phi)$ very useful, 
particularly in 
 studies of spontaneous symmetry breaking (SSB). The effective potential 
determines the one particle irreducible (1PI) vertices~\cite {1} at zero
momenta and reflects any non-trivial dynamics. It is also widely used to 
study radiative 
corrections in quantum field theories~\cite {3}.
Truncating the loop expansion of the effective potential often gives it a 
complex and nonconvex character, in spite of the fact that on general 
grounds the 
effective potential must be real and of convex character~\cite {5}. 
It has been pointed 
out that the loop expansion for the effective potential fails for the 
fields in just those regions where 
 the classical potential is nonconvex; the most familiar case corresponds to
 a double-well potential~\cite {6}. Therefore, it is important to carry out 
nonperturbative studies which can be used even where the loop expansion is 
not applicable. One
 convenient nonperturbative
approach is to employ a discrete version 
of the theory, i.e, lattice field theory. Lattice field theories have an 
ultraviolet (UV) regulator (the
lattice spacing) and and an infrared (IR) cut-off (the lattice size) and 
are conveniently studied using Monte Carlo (MC) methods.

The model used in our study is the $\lambda \phi^4_{1+1}$ model. The Higgs 
mechanism is based on a more 
elaborate version of  such a model and is usually discussed at the tree
 level. A fully nonperturbative treatment of the Higgs 
model would be of considerable interest, but is not discussed further here.

In this work we investigate  the effective potential for 
$\lambda \phi^4_{1+1}$ using three different methods:\newline
The variation of the source method (VSM) ~\cite{callaway1} and two types of 
constraint effective potential, which we denote as CEPI and CEPII,
(see refs.~\cite{6} and ~\cite{8} respectively).

 We 
 point out the advantages and disadvantages of each method and their 
accuracy. Some  suggestions for improving the methods are also put to
 the test. In Sec.~2 we briefly summarise the model to be studied. 
In Sec.~3 we review the above methods of calculation of  
the effective potential. In Sec.~4 we perform the calculations for both the  
symmetric and the spontaneous symmetry breaking cases in the weak coupling 
regime and we compare our results with those obtained from
  lattice perturbation theory. We also perform calculations in the strong 
coupling regime and compare these with the strong coupling expansion.
\section {The $\lambda\phi^4$ model}
We start with the action of a single component 
 $\lambda\phi^4$ theory in d-dimensions in Euclidean space in the presence of  
a source $J$, (in units where $ \hbar =c= 1$)
$$
S[\phi,J]= \int d^dx \ {1\over 2}(\del_{\mu}\phi)^2+{1\over 2} 
m^2\phi^2+{\lambda\over 4!}\phi^4-J\phi.
$$
A discrete lattice version of the action can be written as
\be
S[\wh \phi,\wh J]= \left[ {1\over 2}\sum_{n,\mu} (\hp_{n,\mu}-\hp_n)^2+
{1 \over 2}\sum_n\widehat m^2\hp_n^2+\sum_n{\widehat\lambda \over 4!}
\hp_n^4-\sum_n \widehat J_n\hp_n \right] \label {action},
\ee
where  we have defined the dimensionless quantities 
$\hp \equiv a^{(d/2)-1}\phi$, $\widehat m\equiv m a$ and 
$\widehat \lambda\equiv \lambda a^{4-d}$ and
 $\widehat J\equiv a^{(d/2)+1}J$. In addition $n\equiv (n_1,\dots,n_d)$ is a 
$d$-dimensional vector labeling the lattice sites and $\mu$ is a 
unit vector in the temporal or spatial direction. The sum over $\mu$ is 
over the $d$ Euclidean directions. We also have denoted the field on the 
neighboring site of $n$ in the direction of $\mu$ by $\wh \phi_{n,\mu}$.
 Henceforth we drop the hat from the dimensionless field variables
and sources for brevity unless
it is necessary to avoid confusion. We also impose the 
 appropriate periodic boundary condition on fields:
\be
\phi_{n+ \wh N_{\mu}}=\phi_n \ \ \ \rm {for\  all} \ \ {\mu}\ ,
\ee
where $\widehat N_{\mu}=(0,\dots,N_{\mu},\dots,0)$, is a $d$-dimensional 
vector with
$N_{\mu}$ being the number of lattice sites in the direction $\mu$.

The $\phi^4$ theory is known to exist in two phases, one where the 
reflection symmetry $\phi \rightarrow -\phi$ is spontaneously broken and the 
other where it is not. The symmetric phase with  $\mean{\phi}=0$, is
 separated from
 the broken symmetry phase with $\mean{\phi}\neq 0$ by a line of second 
order phase
transitions where $\widehat m$ and $\widehat\lambda$ assume the critical 
values $\widehat m_c$ and $\widehat \lambda_c$.

 For the action $S[\phi]$ on the lattice the generating functional for the 
correlation functions is defined as: 
\be
Z[J]=  {\int [d\phi] e ^{- S[\phi,J]}\over \int [d\phi] e^{- S[\phi]}},
\ee
such that $Z[0]=1$.
>From $Z[J]$ one can define the connected Green's functions as: 
\be
G(n_1,\dots,n_j)_c= {\del \over \del J_{n_1}}\dots {\del \over \del J_{n_j}}
\left. W[J]
\right|_{J=0},
\ee
where 
\be
W[J]= \ln Z[J].
\ee

\section {The lattice effective potential}
Consider a lattice lagrangian density on a $d$-dimensional cubic lattice with 
the total number of lattice sites $N^d$,
\be
{\cal L}_n= \sum_{\mu}{1\over 2}\left(\phi_{n,{\mu}}-\phi_n\right )^2+ 
V(\phi_n) \ .
\ee
 The classical  vacuum ( ground state) is at the minimum  of $V(\phi)$.
 The vacuum expectation value $\mean{\phi}$ of the quantum field is not 
necessarily identical to the classical vacuum.
The vacuum expectation value of the field in the presence of an external 
source,
$J(x)$ is given by 
\be
\phi_{\mbox{\tiny c} n}[J]\equiv {\del W[J] \over \del J_n}  \label {a1} \ .
\ee
The vacuum expectation value $\mean{\phi}$ is the limit of 
$\phi_{\mbox{\tiny c}\it n}$ as $J 
\rightarrow 0$. Hence we can ask for  what value of $J$  can one obtain a 
given $\phi_c$.
One can choose to treat $\phi_c$ as the independent variable instead of $J$ and
define the ``effective action'' $\Gamma[\phi_{\mbox{\tiny c}}\rm]$ by a 
Legendre transformation:
\be
\Gamma[\phi_c] =  \sum_n  \phi_{\mbox{\tiny c}\it n} J_n - W[J] ,\ \label 
{legendtr}
\ee
 where $\phi_c$ is defined in Eq.~\eqn {a1}.  
It is easy to verify that
\be
 J_n[\phi_c]\equiv {\del \Gamma[\phi_{c}\rm] \over \del \phi_{\mbox{\tiny c}
\it n}\rm} \label{a3} \ .
\ee
In the case  $J=0$, by translational invariance it follows that $\phi_c\rm$
 must become constant (i.e, independent of the label $n$). Hence the vacuum 
expectation value is given by  $\mean{\phi}$ and satisfies
\be
\left.{d\Gamma[\phi_c] \over d\phi_c}\right|_{\phi_c=\mean{\phi}}=0 \label{a4}
\ .
\ee
Similarly for any constant $J$ we must have $\phi_c=\ov \phi$ also constant.
Define the effective potential, $U(\ov \phi)$,  by
\be
\Gamma[\overline \phi]= N^d U(\overline\phi)  \label {a5} \  .
\ee
\newline
The Fourier trasform on a finite, discrete lattice is defined by 
\be
\wt \phi_k \equiv \sum_n e ^{2\pi i n.\hk/N}\phi_{n},
\ee
where $\hk\equiv \hk_1,\dots,\hk_d$ is a $d$ dimensional vector with 
$ (-N/2)<\hk_n\leq N/2$. (we assume $N$ is even from this point) and where 
$n.\hk\equiv n_1\hk_1+\dots+ n_d \hk_d$. The coordinate-space and
 momentum-space $\delta$-functions are 
\be
\delta_{m,n}= {1 \over N^d}\sum_{\hk} e^{-2\pi i(n-m).\hk/N} \ \ \ \ \ \ \ \ \ 
\ \  \ 
\delta_{\hk,\hq}= {1 \over N^d}\sum_{n} e^{-2\pi i(\hk-\hq).n/N}
\ee
respectively. The inverse Fourier transform is 
$$
\phi_{n}= {1 \over N^d} \sum_{\hk} e ^{-2\pi in . \hk/N}\wt \phi_{\hk}
$$
Note that we have used the asymmetric normalisation of the Fourier transform 
and 
its inverse as is usual in the field theory in the continuum. 
The effective action is the generator of proper (i.e, one-particle irreducible)
Green's functions and in particular we can Taylor expand the effective action 
to give
\be 
\Gamma[\phi_c]=\sum_{M=0}^{\infty}{1\over M!}\sum_{n_1,\dots,n_M }
\Gamma^{(M)}(n_1,\dots,n_M)\phi_{cn_1}\dots \phi_{cn_M}.
\ee
Here $\Gamma^{(M)}(n_1,\dots,n_M) $ is the proper $M$-point Green's functions 
in 
presence of the source $J_n$
\be
{\del^M\Gamma[\phi] \over \del\phi_{cn_1}\dots \del{\phi_{cn_M}}}= \Gamma^{(M)}
(n_1,\dots,n_M).
\ee
 In terms of its Fourier transforms we have 
\be
\Gamma[\phi_c]= \sum_{M=0}^{\infty} {1 \over M!}{1 \over N^{Md}}\sum_{\hk_1,
\dots,\hk_M} \wt \Gamma^{(M)}(\hk_1,\dots \hk_M)\wt \phi_{-\hk_1}\dots 
\wt \phi_{-\hk_M} \label {gfc},
\ee
where here $\tilde \phi$ is the Fourier transform of $\phi_c$ 

The vacuum proper Green's functions are obtained by setting $J=0$. If the 
source is a constant ($J_n=J $ for all $n$), then the translational 
invariance is 
restored 
and we can factor out an overall normalisation and a $\delta$-function to 
define
\be
\wt \Gamma^{(M)}(\hk_1,\dots,\hk_M)\equiv N^d\delta_{0,\hk_1+\dots+\hk_M}
\wt \Gamma^{(M)}_c(\hk_1, \dots, \hk_M) \label {3.12}
\ee
Where in the limit $J\rightarrow 0$ we recognise that
 $\wt \Gamma^{(M)}_c(\hk_1,\dots,\hk_M)$ is the dimensionless, lattice 
equivalent of the proper $M$-point Green's function in momentum space.
For constant $J$ we also have $\phi_{cn}\rightarrow \bp$ and hence $\wt \phi=
N^d\delta_{\hk,0}\bp$ and Eq. \eqn {gfc} gives
\be
\Gamma(\ov \phi)=N^d\sum_{M=0} {1 \over M!}\wt \Gamma^{(M)}_c(0)\ov \phi^M ,
\ee
where 
\be
\wt\Gamma^{(M)}_c(0)\equiv \wt\Gamma^{(M)}_c(0,\dots,0) \ .
\ee
Comparing Eq.\ \eqn{a5} with Eq.~\eqn {gfc} gives
\be
U(\overline\phi)= \sum_{M=0}^{\infty}{1 \over M!}
 \wt\Gamma^{(M)}_c(0) \bp^M \label{ef1} \ .
\ee
It immediately follows that
\be
\left.{d^MU(\overline\phi) \over d\overline\phi^M}\right|_{\overline\phi
=\mean{\phi}=0}=\wt\Gamma^{(M)}_c(0) \ ,
\ee
where here it is understood that we are working in the unbroken symmetry 
sector, $\mean{\phi}=0$.
In the unbroken sector we see from \ Eq. \eqn {ef1} that the dimensionless, 
proper Greens functions with vanishing momenta 
can be easily obtained from the effective potential, $U(\phi)$, by  
differentiation.
 We see that $\bp$ minimises $U(\bp)$ and in the limit $J\rightarrow 0$ 
the minimum $\bp\rightarrow \mean{\phi}$.
Also note that Eq.~\eqn {a3} gives an expansion of $J$ in terms of the 
$\ov \phi$'s and $\Gamma(0)$'s
\be
J(\overline \phi)=N^d \sum_{M=1}^{\infty}{1 \over (M-1)!}\wt\Gamma^{(M)}_c(0)
  \ \ \ov \phi^{^{M-1}} \label {jsource} \ .
\ee
In the broken symmetry sector, $\mean {\phi}\neq 0$, it is more appropriate 
to use the shifted field
\be
\chi(x)\equiv\phi(x)-\mean{\phi} \ .
\ee
The one-particle irreducible (1PI)  vertex functions  $\Gamma^{(M)}_{(s)}$ 
are linear 
combination of the $\Gamma^{(M)}$'s, and can be obtained from the shifted 
version of Eq.~\eqn {ef1} 
\be
U(\bp)\equiv U_{(s)}(\ov \chi)=U_{(s)}(\ov\phi - \mean{\phi})= 
\sum_{M=0}^{\infty} {[\overline\phi-\mean{\phi}]^M \over
M!} \ \wt\Gamma_{(s)}^{(M)}(0) \ .
\ee
As is usually done in lattice field theory studies we renormalise at the 
renormalisation point where all external momenta of the Greens functions 
vanish. 
The renormalised quantities can be obtained directly from the effective 
potential.
 For example in the $\lambda\phi^4$ theory we have:
\be
\left.{dU(\overline\phi) \over d\overline\phi}\right|_{\overline\phi
=\mean{\phi}}= 0 \ \ \ \ \ \ \ \ \ \ \ \ \ \ \ \ \ \ 
\ee
\be
Z \left.{d^2U(\overline\phi) \over d^2\overline\phi}\right|_{\overline\phi
=\mean{\phi}}=
Z \wt\Gamma^{(2)}_c(0)= \wt \Gamma_r^{(2)}= \wh m_r^2
\ee
\be
Z^2 \left.{d^4U(\overline\phi) \over d^4\overline\phi}\right|_{\overline\phi
=\mean{\phi}}=
Z^2\wt\Gamma^{(4)}_c(0)=\wt \Gamma_r^{(4)}=\wh\lambda_r \label {recond} \ .
\ee
where $\sqrt Z$ is the field wavefunction renormalisation constant ($\phi_r=
\sqrt Z \phi$).
>From the first two conditions above and requiring $\wh m_r^2 \geq 0$ it 
follows that $\mean{\phi}$ is
at the minimum of $U(\overline\phi)$. Also note 
that $\wh m_r$ and $\wh \lambda_r$ defined 
as above are not the physical mass and coupling, which are defined in the 
pole
of the propagator in the complex energy plane and the on shell four-point 
function, respectively. However in the scaling region (close to the critical 
line) 
these values are a good approximation to the physical mass and 
coupling~\cite {7}. 
\section {The MC effective potential.}
In this section we will examine three MC methods for calculation of the 
lattice
 effective potential. The renomalised coupling constants obtained by these 
 methods are compared with analytical results. From this point on we work 
exclusively in two dimensions (d=2). 
\subsection {The variation of source method (VSM)}
Eq.~\eqn {jsource} suggests that in the Monte Carlo calculation one can 
calculate the mean value of the fields, $\overline\phi$, for different 
values of 
the source and as 
a result one obtains $\overline \phi$ as a
function of $J$. This function can then be inverted to obtain $ J$ as a 
function of $\ov \phi$, i.e., $J(\overline \phi)$. Then  using 
Eq.~\eqn {jsource} we see that the derivatives of $J$ with respect to 
$\overline \phi$ 
would give the proper Green's functions at zero momentum.
>From Eq.~\eqn {a1} one also concludes that $\bp_J$ is antisymmetric
in $J$. That is
\be
\bp_J=-\bp_{-J} \ .
\ee
 Fig.~1 shows $J(\bp)$  as a function of $\bp$ for the 
 symmetric case (Fig.~1a) and the broken symmetry case (Fig.1b). Note that for
 the broken symmetry case, $\bp_J$ as a function of $J$ is discontinuous
 and so the relation in Eq.~\eqn {jsource} can not be inverted for all 
$\bp(J)$.
 Whenever it is possible Eq.~\eqn{jsource} has to be inverted to obtain 
the source $J$ as a function of $\bp$. 
Then the derivatives of $J$ with respect to  $\overline \phi$ would give the 
vertex functions at zero momenta and 
consequently the renormalised masses and couplings can be calculated. 

The mean value of the field in the presence of a source has a small 
 statistical error. This is expected since it is an analog to the reduction of
 fluctuations of a spin system in the presence of an external magnetic field. 
 As the source becomes smaller the fluctuations become larger. Thus one needs 
 to perform the calculations for large enough sources that the error is small
 and then extrapolate the 
 results to $J=0$.

This method will be referred to as the variation of source method (VSM) and 
has a number of advantages. The vacuum expectation values of the 
 field $\ov\phi(J)$ are the simplest quantities to compute on the lattice 
and their $J$-dependence can be exploited to get the first derivative of the 
effective potential.
Since the source effectively causes the boson field to become more massive, 
the finite size 
effects generated by the lattice become exponentially small provided that 
the lattice is large enough. Since the 
data become noisy  for small values of $J$ we need to restrict 
the analysis to a safe region of $J$, which can introduce some errors in 
the results through uncertainties in the extrapolation.\newline 
\subsection { The Constraint Effective Potential (CEP I)}
 In the last section the effective action and the effective potential 
$U(\overline\phi)$ were defined
 through introduction of a source $J$. There is a different method which does 
 not require such a dynamical symmetry breaking source. 
The constraint effective potential was first introduced by Fukuda and
 Kyriakoloulos~\cite {8} as an alternative way of obtaining 
 the explicit expression for the effective potential. It was further analyzed 
by O'Raifertaigh, Wipf and Yoneyama~\cite {9}. In this approach one  obtains 
an explicit expression for the effective 
 potential, without introducing external sources, but instead through the 
introduction of a 
$\delta$-function in the functional integral. In the constraint effective
potential approach one first defines $ \wt U(\bp)\equiv U(N^2,\overline\phi)$ 
as
\be
e^{-N^2 \wt U(\overline\phi)}=\int [d\phi] \delta\left({1 \over N^2}\sum_n
\phi_n- \overline\phi\right)e^{-S[\phi]} \label {a8} \,
\ee
and then uses the fact that as $N^2\rightarrow \infty$ we have $\wt U(\bp)
\rightarrow U(\bp)$ and the effective potential is recovered.
{W

It is easiest to demonstrate this result in Minkowski space, where
 Eq.~\eqn {a8} becomes
\be
e^{-iN^2\wt U(\overline \phi)}=\int[d\phi]\delta({1\over N^2}\sum_n \phi
_n-\overline \phi) e^{iS[\phi]}. \label {a9}
\ee
We can replace the $\delta$-function in Eq.~\eqn{a9} by its integral
representation to obtain (up to an irrelevant constant)
\ba
e^{-iN^2\wt U(\overline \phi)}&=&C\int dJ\int [d\phi]e^{i\int dx[{\cal L}+
J\phi]-iN^2\bp J}\nn
&=& C'\int dJ \ e^{i(W[J]-N^2J\bp)} \label {mink}
\ea
Note that in the integrand of Eq.~\eqn{mink} we have $\bp$
 fixed and $J$ arbitrary. In the limit, $N^2 \rightarrow \infty$ the dominate 
contribution to the integral 
comes from the stationary point of the integral which is the value of
 $J$ at which $d W[J]/dJ=\bp$. Recall that
 $\Gamma(\bp)=\left.(J\bp-W[J])\right|_{\bp=d W[J]/dJ}$, from which we see 
that up to an irrelevant overall constant
\begin{equation}
e^{-iN^2\wt U(\overline \phi)} \rightarrow  e^{-i\Gamma(\overline \phi)}
=e^{-iN^2U(\bp)}  \ \ \ \ {\rm as } \ N^2 \rightarrow \infty,
\end{equation}
as claimed.

We can also arrive at this result directly in Euclidean space by multiplying 
both sides of Eq.~\eqn {a8} by 
$e^{N^2J\ov \phi} $ with $J$ arbitrary and then integrating over
$\ov \phi$ to obtain
\be
\int d\bp \  e^{-N^2[\wt U(\ov \phi)-J\bp]}\ =\int [d\phi] \ e^{-S[\phi]+
J\sum_n \phi_n}.  \label {F1}
\ee
As $N^2\rightarrow \infty$ the left hand side of Eq.~\eqn {F1} becomes 
entirely dominated by the stationary point of the one-dimensional $\bp$
integration given by $d\wt U(\bp)/d\bp=J$, while the right hand side
is recognised as $e^{W[J]}$ for a constant source, $J$. Hence up to
an irrelevant overall constant we find
\be
e^{-N^2[\wt U(\bp)-J\bp]}\rightarrow e^{W[J]} \ \ \ \  \ \ {\rm as} \   \ 
N^2\rightarrow \infty,
\ee
and so find that (up to a constant)
\be e^{-N^2\wt U(\bp)}\rightarrow e^{W[J]-N^2J\bp}= e^{-N^2U(\bp)} \ \ \ \ 
{\rm as}\  \ N^2\rightarrow \infty,
\ee
as required.

 It is important to note that the $e^{-N^2\wt U(\overline\phi)}$ relates 
to similar definitions in statistical mechanics and spin 
systems~\cite {10} and that 
\be 
P(\overline \phi)= {e^{-N^2 \wt U(\overline \phi)} \over \int d\overline 
\phi \  e^{-N^2 \wt U(\overline \phi)}}   \label {probdis} \ .
\ee
can be interpreted as the probability density for the system
to be in a state of ``magnetization'', $\overline\phi$. Then it can be seen 
that the 
probability for a state whose average field is not a minimum of 
$\wt U(\overline\phi)$ then decreases as $N^2 \rightarrow \infty$.
 
This suggests that one needs to study the probability distribution of the 
order 
parameter $\overline \phi$. Using a Monte Carlo algorithm one generates a 
Boltzman 
ensemble of configurations, $\{\phi\}$, weighted by $e^{-S[\phi]}$. Let 
${d{\cal N}}$ be the number of configurations 
with average field values in an interval $d\overline \phi$ about $\overline
\phi$. Then
\be
d{\cal {N}}(\overline \phi) = C  e^{-N^2   \wt U(\overline \phi)}d\overline 
\phi \ ,
\ee  
with $C$ some constant. Then one can write
\be
\wt U(\bar\phi)= -{1 \over N^2} \ln {d{\cal N}(\bar\phi) \over d\bp} \label 
{prob2},
\ee
up to an irrelevant additive constant.
Eq.~\eqn {probdis} suggests that one can generate a large number of
 configurations weighted by $e^{-S[\phi]}$, calculate $\bp$ for each 
configuration and construct a 
normalised histogram. The histogram can be fitted to Eq.~\eqn{prob2}.
 The most probable average field values are 
near the minimum of the effective potential. In order to determine 
$\wt U(\overline\phi)$ 
away from its minimum,{\it i.e.}, to sample a range of relatively 
improbable values of $\overline\phi$, one can introduce a small source. 
Then a simple generalisation of Eq.~\eqn {probdis} allows a nonzero external
 source ~\cite {Tyson}
\be
\wt U(\bar\phi)-J\bp= -{1 \over N^2} \ln {d{\cal N}(\bar\phi) \over d\bp}
\label {anzats} .
\ee  
Thus one can check whether such an ansatz gives a good approximation for the 
effective potential, and so construct the effective potential by performing 
a simultaneous
fit of several histograms corresponding to different values of $J$. By the
expression
``simultaneous fit'', we mean that the chi-squared values corresponding to 
each $J$ are
 summed and this sum is then minimised.
This method can be applied easily on the lattice. Note
 that in Eq.~\eqn {probdis} we have assumed that for sufficiently large 
$N^2$ the 
 finite volume effects on $\wt U(\overline\phi)$ can be neglected, {\it i.e.}, 
that the
lattice volume is sufficiently large. The constraint effective potential 
method summarised in Eq.~\eqn {anzats} will be refered to as CEPI.

\subsection { The Constraint Effective Potential (CEP II)}
Now return to Eq.~\eqn {a8} and perform a shift of field,
$\phi(x)\rightarrow \phi(x)+\overline\phi$. Since the measure is 
translationally invariant we obtain 
\be
e ^{-N^2 \wt U(\overline\phi)}= \int [d\phi] \ \delta ({1\over N^2}
\sum_n\phi_n) \ \ e^{-S[\phi+\bp]}
\ee
Taking the derivative with respect to $\overline\phi$ we get 
\be
{d\wt U(\overline\phi) \over d\overline\phi} e^{-N^2\wt U(\overline\phi)}= 
{1 \over N^2} 
\int  [d\phi] \ \delta ({\sum\phi_n\over N^2}) {dS(\phi+\bp) \over 
d\overline\phi}\ e^{-S[\phi+\bp]}.
\ee  
Only the potential part of the action is affected by the shift of field since 
$\bp$ is constant and so $dS/d\bp=N^2dV/d\bp$. Using this fact and shifting
 the field back to its original form then gives
\be
{d\wt U(\overline\phi) \over d\overline\phi }= \left\langle {dV(\phi) \over 
d\phi}\right \rangle_{\overline\phi},
\ee
 where we have introduced the shorthand notation 
\be 
\mean{O(\phi)}_{\overline\phi}\equiv (e^{N^2 \wt U(\overline\phi)})\int \  
[d\phi] \ \ 
\delta({1\over N^2}\sum_n \phi_n \ -\overline\phi)\ \ O(\phi) \ e^{-S[\phi]}.
\ee
 In the $\lambda\phi^4$ theory being considered here we find
\be
{d\wt U(\overline\phi) \over d\overline\phi} = \widehat m^2\overline\phi+ 
{\widehat\lambda  \over 6}\mean{\phi^3}
_{\bp}. \label {firstd}
\ee
Expressions for some of the higher derivatives of $U(\bp)$ 
 are given in Appendix A. These equations are very useful in the Monte Carlo 
(MC) calculations since they relate the  derivatives of the effective 
potential (and consequently the 
zero momentum vertex function) to the averages of quantities that can be 
calculated directly from the lattice. This method will be refered to as CEPII.

 There are two ways of 
calculating the renormalised quantities using CEP II. The first one applies 
the constraint on the lattice, fixing
 $\overline\phi$, then calculates $\mean{\phi^3}$, and finally uses 
Eq.\ \eqn{firstd} to obtain the 
first derivative of the effective potential. Higher derivative are evaluated 
from fitting a curve to the $d\wt U/d\bp$ versus $\bp$ results. This has some 
similarities with
the variation of source method, however there is a difference between these 
two methods. In VSM one sets the source $J$ to constant and 
$\mean {\phi}=\bp$ up to fluctuations due to finite $N^2$, whereas in CEPII we 
have $\mean {\phi}=\bp$ exactly by construction.

 In the broken symmetry sector there is another difference 
between this method and VSM in the broken sector. When using VSM we are not 
able
to obtain 
 any value of $\mean{\phi}$ in the region between the two minima, whereas both
CEP methods are suitable for probing this region. One can always 
 fix $\bp$ to any value including the values between the two minima to get the 
 full shape of $J(\bp)$ (see Fig.2).  
However as far as the practical calculation of renormalised quantities is 
concerned, 
this method is almost equivalent to VSM and so from here on we disregard 
this approach.

The second approach to CEPII is through the equations shown in Appendix A 
and is
more direct. These equations relate the derivative
of the effective potential to the averages of some derivatives of the 
classical potential. All these averages should be taken in the presence of
 the constraint which fixes $\mean{\phi}=\bp$.

Imposing a constraint on a Monte Carlo algorithm is relatively simple. One 
starts with 
a configuration with the field average  being $\bp$ \cite {8}. Each time a
 site is updated by a value $\delta$ (say) such that 
\be
\phi_i' = \phi_i +\delta \ ,
\ee
then some arbitrary site $k$ must be updated simultaneously such that 
$\phi_k'=\phi_k -\delta$.
This procedure is carried out for all the sites, which completes a sweep. 
For a large enough lattice imposing such a constraint is not expected to
 violate the ergodicity of the MC algorithm.  

The advantage of this method over the VSM is that one does not need
 to run a Monte Carlo routine several times with different sources, and no 
curve fitting is required. 
 One disadvantage of this method is that for calculation of the 
 renormalised coupling one needs to add and subtract many average terms as 
has been shown in Appendix A.
Although the statistical errors might be small
 for each term, the 
overall errors contributing to the renormalised coupling can be large. However,
the renormalised mass in the symmetric phase of the $\lambda\phi^4$
theory obtained using this method is very accurate. 

 We also would like to comment on Fig. 2. It has been shown by a very general 
argument that $U''(\bp)\geq0$ for all $\bp$~\cite {5}, (primes denote 
differentiation with respect to $\bp$). This general property is known as 
the `` convexity'' of
 the effective potential. Looking at Fig. 2 it is  clear  
that this condition is violated for $\wt U(\bp)$. This can be understood 
by noting that convexity holds only in the thermodynamical limit, {\it i.e.}, 
$N^2\rightarrow \infty$.

To conclude this section it should also be mentioned that the proper vertex
 functions can be obtained directly using the standard Monte Carlo (MC) method.
 For example for the $\lambda \phi^4$ four point vertex function one obtains
\be
\wt \Gamma^{(4)}_c(0)=- {\mean{\wt \phi^{^{^4}}}_c-3\mean 
{\wt \phi^{^{^2}}}_c^{^{2}} \over
\mean{\wt \phi^{^{^{^2}}}}_c} \label {4vertex} \ .
\ee
Here, for example, $\mean{\wt\phi^4}_c$ is the connected part of vacuum 
expectation value of fourth power of the Fourier transform of the field at 
zero 
momentum.
As we will show, in the weak coupling regime this method suffers from very 
noisy signals giving rise 
to large statistical errors. The errors are due to the large fluctuations of
 correlation functions in this regime as well as the subtraction of the 
disconnected 
pieces. 
 However in the strong coupling regime this 
method gives a relatively good approximation for $\wt\Gamma^{(4)}_c(0)$ and 
the statistical errors
 are reasonably small~\cite {11}. However the higher order vertex functions 
calculated with this approach can be very noisy even in the strong coupling 
regime, primarily due to subtractions of noisy disconnected pieces. 

\section {The numerical results}
In this section we present our results for the calculation of renormalised 
coupling, 
$\lambda_r$, in two dimensions. It includes the symmetric and 
 broken symmetry sector in the weak coupling regime as well as the strong 
coupling regime. In the case of the weak coupling regime the results are 
compared with 2-loop results and the direct calculation of $\lambda_r$ using 
the standard MC method in Eq.~\eqn {4vertex}. In the strong coupling regime we 
also compared the results of each 
 method with the strong coupling expansion results. 
 The details of the numerical simulation are included at the end of this
 article.\newline
\subsubsection* {Case 1: The symmetric sector in the weak coupling regime 
(WCR)}
{\it{The variation of the source method (VSM):}}\newline
 Here we study the model in the symmetric sector where $\mean {\phi}
=0$. As we will see, all methods presented in this paper require the calculation of 
renormalised mass $\wh m_r$, and the wavefunction renormalisation constant, 
$Z$.
In general, the boson propagator extracted from the lattice has the form 
\be
\wt G{(\hk)}=  {Z \over { \hk^2 + \widehat m^2_r(\hk^2)}}  \ ,
\ee
where $\wh m_r\equiv \wh m_r(\wh m_r^2)$ is the mass-pole of the scalar 
particle, i.e., the
 renomalised mass. 
 In particular at zero momentum 
\be
\wt G(\hk=0) =  {Z \over {  \widehat m^2_r}} \label{abas3} \ ,
\ee
where we make the standard approximation that $\wh m^2_r\approx
\wh m_r^2(0)$.
The renormalised mass, $\widehat m_r$, is then given as the reciprocal of 
$\widehat\zeta$,
 the lattice correlation length
\be
\widehat \zeta^2 = {1 \over {\widehat m^2_r}} = \left. {1 \over {\wt G(\hk)}}
 {{d\wt G(\hk)\over d\hk^2}} \right|_{\hk=0} \label {abas4} \ .
\ee
Taking into account the translational invariance of the correlation functions,
 one
can choose to approximate the momentum derivative in Eq.~\eqn{abas4} by the
variation  of 
$\wt G(\hk)$ across one lattice spacing and in one direction to obtain
\be
\widehat \zeta^2 = {N \over {2\pi}}\left [ {\mean{\gamma^2}_c -
\mean{\alpha^2}_c-\mean{\beta^2}_c}\over
 { \mean{\alpha^2}_c+\mean{\beta^2}_c}\right] \label {masscal} \ ,
\ee
with 
$$
\alpha = \sum_{n=1}^N \sum_{m=1}^N \ \phi_{n,m} \  \cos\left[{{2\pi}\over N} 
(n-{N\over 2})\right] 
$$
$$
 \beta = \sum_{n=1}^N \sum_{m=1}^N \ \phi_{n,m} \ \sin\left[({{2\pi}\over N} 
(n-{N\over 2}))\right]
$$
\be
 \gamma=\sum_{n=1}^N \sum_{m=1}^N \ \  \phi_{n,m} ,
\ee
where here $n,m$ label the temporal and spatial coordinates for the
field $\phi$ respectively.

 There are two different ways
 of calculating $Z$. One is to use Eq.~\eqn {abas3} and the fact that 
$\wt G(0)=N^2\mean{\wt\phi^2}$ to calculate $Z$.
The second way of calculating $Z$ comes from combining Eqs.~\eqn {abas3} 
and~\eqn {recond}
which gives
\be
\left.{d^2U(\bp)\over d\bp^2}\right|_{\bp=\mean{\phi} }= {1 \over \wt G(0)}
\label {cond} \ .
\ee
Thus $\wt G(0)$ can be directly calculated from the fit and the calculation of
 $Z$ follows as before. An accurate calculation of $\wh m_r$ is crucial for 
both 
methods. We found that in the weak coupling regime, the second method was 
more precise.
 We compared our results with the 2-loop lattice perturbation theory 
calculations (LPT) of the renormalised parameters.
 This means that finite size effects may be present in our comparisons at 
some level. 
The comparison is shown in Fig. 3. The values for $\lambda_r$ seem to be
 accurate 
 even in the very weak coupling regime. In this regime the effective
potential results are in good agreement with the lattice perturbation 
 calculations. The MC results begin to deviate from the perturbative 
calculations 
 as $\widehat\lambda_r$ increases. This is expected since a loop expansion 
in $\lambda\phi^4$ theory is an expansion in $\widehat\lambda_r$ and as this is
increased the contribution from higher loops becomes more 
significant.  

The VMS can be expensive in CPU time but the cost can be reduced to 
some extent. For a value of $J$ it is possible to calculate $D^n(\phi)\equiv 
d{\wh\phi}^n_J/d^nJ=N^2\mean{\wt \phi^n}_c$ during the calculation of 
$\phi_J$, for 
each value of $J$.
 From these derivatives one can expand 
$\bp_J$ around $J$ and then use a curve fitting routine to calculate 
$\lambda_r$, as we did before. 
The statistical errors can become larger for the higher derivatives because 
of the
 subtraction of the disconnected pieces of $D^n(\phi)$. 
In Table~1 we have shown a 
comparison of our previous results for $\bp_J$ and results obtained by 
expansion around three $J$ values, namely, $J=0.075,0.25,0.4$ for 
$\widehat m^2=0.1,
\widehat \lambda=0.1$. We see that the calculated values of $\bp$ are 
reasonably
close to the previous results. However, the price for reduced computational
 time is a slight increase in uncertainties.

We have also calculated $\widehat\lambda_r$ for $\widehat\lambda=0.055, 
\widehat m^2=0.1$ using Eq. \eqn {4vertex} and the result is included in 
Fig. 3. 
The statistical errors are extremely large and it suggests that the 
calculation of the 4-point vertex function in this region is impractical with 
this method.

{\it The constraint effective potential method I (CEPI):} 

This method is the  easiest to implement. We 
generated the Boltzman ensemble of independent configurations. For every 
configuration we measured $\overline\phi= ({1/N^2})\sum_i\phi_i$
 and computed the histograms for the probability density
$P(\overline\phi)$ for several values of $J$. We also noticed that the
 anzatz of Eq.~\eqn {anzats} 
only worked well for very small $J$ in this region. We did a simultaneous fit 
to Eq.~\eqn{probdis} of a few histograms corresponding to $J=0$ and small
$J$'s  using a three-parameter anzatz for $\wt U(\overline\phi)$ of the form
\be
\wt U(\bp)= a_1\bp^2+a_2\bp^4+a_3\bp^6 \label{param}.
\ee
 Although there was no systematic discrepancy between the data 
and the fit, the statistical errors were very large.
We unsuccessfully tried  more histograms and higher powers
of $\bp$ in the fit. The statistical errors remained large and we concluded 
that 
even a reasonable estimate of renormalised parameters in this region was not
feasible with this method.

{\it The constraint effective potential method II (CEPII):}

  In this method Eqs.~(\ref{app4}) and (\ref{app8}) 
can be used for calculations of 
$\wh m_r$ and $\wh\lambda_r$ respectively. All averages shown in these 
equations are
to be taken with the constraint of $\overline\phi=0$. 
 Although the statistical errors for each term are small,
the overall error can be large. 
However in the symmetric case in in the weak coupling regime most of the  
terms either vanish at  
$\overline\phi=0$ or are small enough to be neglected. For example for 
$\wh m_r$ only three terms need to be considered. But the computation of 
$\wh\lambda_r$ suffers from 
larger cumulative errors. 

 The results are compared with the VSM results and are shown in Table~2. We 
also 
compared the calculation of renormalised mass using Eq.~\eqn {masscal} with 
the 
CEPII calculations in Table ~3. The comparison indicates that in this sector 
the
CEPII method can provide an accurate calculation of the renormalised vertex 
functions.

\subsubsection* {Case 2: The broken symmetry sector}
In this section we consider the calculation of the renormalised mass and 
renormalised coupling in the broken sector, $\mean{\phi}\ne 0$, in the  
weak coupling regime. The VSM procedure is exactly the same
 as for the symmetric sector. For fixed $\widehat m^2=-0.1$
 and $0 < \widehat\lambda \leq 0.17$ we calculated $\widehat m_r$ and 
$\widehat\lambda_r$
 for different values of $\widehat \lambda$.
 The error on  $\wt G(0)$ is larger than the symmetric case 
due to the subtraction of the disconnected pieces. Thus we used 
Eq.~\eqn{cond} to calculate 
  $\wt G(0)$ and subsequently extracted $Z$ as previously discussed.

In order to calculate the renormalised quantities using lattice perturbation 
theory we followed 
 the standard approach to treating the broken sector. That is, in the bare 
Lagrangian
we shifted the field by the classical value of the vacuum, $\nu= \sqrt {
{-6\widehat m^2 / \widehat\lambda}}$, such that
\be
\chi(x)=\phi(x)+\nu \ .
\ee
After this translation the mean value of the field, $\mean{\chi}$, vanishes 
and 
the perturbative calculation proceeds in the standard manner, keeping in mind 
that a non-symmetric
$\chi^3$ interaction has been generated. In lattice perturbation theory 
one then needs to also consider 
vertex functions with a three point interaction. 

The comparison between the two-loop results and the  results from the VSM 
method are shown in Fig.~4. In applying the CEPII method to the broken 
symmetry 
sector, evaluation of all the
terms in Eq. (app8) is necessary. This renders this method impractical. 
As one might expect from the symmetric sector results the calculation of 
the renormalised parameters using CEPI also suffers from large
noise difficulties and the signal could not be recovered.

\subsubsection* {Case 3 :Strong coupling regime}

 In a weak coupling 
expansion the interactive term is pulled out of the path integral
representation of the partition function as 
a functional operator. That is 
\be 
Z[\wh J]=\exp{\left [{\widehat \lambda \over 4!}\sum_n {\delta^4 \over 
\delta \wh J_n^4}\right]}\int [d\phi] \exp \left[ {-\sum_{n,\mu} {1\over2}
(\phi_n-\phi_{n,\mu})^2
+{1\over 2}\wh m^2\phi_n^2+\wh J_n\phi_n} \right ] \ .
\ee
The remaining functional is Gaussian and can be done exactly. The
partition function can then be written in terms of a power series of 
$\widehat \lambda$ and the standard perturbation theory follows.

The strong coupling expansion was first proposed by the authors of 
Ref.~\cite{strong1}. 
 For this expansion, 
unlike the weak coupling expansion, the kinetic and the mass terms are pulled 
out of the path integral as a functional operator. That is
\be
Z[\wh J]= \exp {\left[{\sum_{m,n} {\delta \over \delta \wh J_n} G^{-1}(n,m)
{\delta \over \delta \wh J_m}}\right]} Z_0[\wh J] \label {stexp} \ ,
\ee
where
\be
Z_0[\wh J]=\int [d\phi] \exp {\left [\sum_n \ {\wh \lambda \over 4!} 
\phi_n^4 +\wh  J_n\phi_n \right ]} \ .
\ee
The remaining functional integral is not Gaussian but can be evaluated as a 
product of ordinary functions on the lattice,
\be
Z_0[\wh J]= {\cal N} \prod_n {F(x) \over F(0)},
\ee
where
\be
F(x)\equiv \int dz e^{-\left [{\widehat\lambda \over 4!} z^4 
+  xz \right ]} \ .
\ee
and ${\cal N}$ is a constant. The function $F(x)$ is a transcendental 
function and can be expanded as a power
series in $x$
\be
F(x)= {1 \over \sqrt(2)} \sum_{n=0}^{\infty} {2^n x^{2n} \over (2n)!}
 \ \Gamma({n\over 2}+{1 \over 4})\ .
\ee
 Using this series expansion one can easily 
expand the both terms in the r.h.s. of Eq. \eqn {stexp} to obtain a power 
series expansion 
for $Z[J]$ which assumes the general form 
\be
Z[\wh J]= {\cal {N'}}\left [1+\sum_{k=1}^{\infty} \widehat\lambda^{-k/2}A_k
[\wh J] \ 
\right].
\ee 
where $A_k[\wh J]$ are integrals over the source function $J$.
Thus the strong coupling expansion is an expansion in powers of 
$\wh \lambda^{-k/ 2}$.
 Bender et al.~\cite {strong1,strong2} obtained a
 series expansion for
$G=  \wh \lambda_r/ \wh  m_r^{4-d}$  of the form
\be
G= y^{-d/2} \sum_{l=0}^L \sum_{n=0}^Na_{nl} \ x^n y^l \label {lamexp} \ ,
\ee
where 
\be
x= y^{-d/2} {\widehat m_r^2 \over \widehat \lambda_r} \ \ \ {\rm {and}}\ \ \  
y=\widehat\zeta
^2= {1\over \widehat m_r} \label {xy} \ .
\ee
For fixed $x$ one has 
\be
G= y^{-d/2} \sum_{l=0}^L \ a_l^{(N)}(x) \ y^l \ ,
\ee
where 
\be
a_l^{(N)}(x)= \sum_{n=0}^N a_{ln}x^n \ .
\ee
This series does not converge  for large correlation lengths. Thus the 
authors of Ref.~\cite {strong2} proposed a scheme to extrapolate the 
expression for 
$\widehat \lambda_r$ to large $y$ assuming that $\widehat \lambda_r$ remains 
finite in the limit
 $y \rightarrow 0$. 
 
Raising Eq.~\eqn {lamexp} to the power of $2L/d$ and expanding to order $L$ 
we find 
\be
G^{2L \over d}= y^{-L} \left (\sum_{l=0}^L a_l^{(N)}(x)
 y^l\right)^{2L \over d}\equiv y^{-L} \sum_{l=0}^L b_l^{(N)}(x)y^l \ .
\ee
We then find  
\be
G= y^{-d/2}\left (\sum_{l=0}^{L}b_l(x)
 y^l\right)^{(d/2L)} \label {extrap} \ ,
\ee
which is equivalent to Eq.~\eqn {lamexp} for small $y$ and approaches 
$[b_L^N(x)]^{d/2l}$ in the 
limit $y \rightarrow \infty $.
In this manner the authors of Ref.~\cite {strong3}  obtained an analytical 
series for Eq.~\eqn {extrap}. 
Since the interesting physics lies in a regime where the correlation length is 
 large, we performed our calculation in this regime. Thus the above 
extrapolation scheme was necessary.

We chose a moderate correlation length $\zeta=3.6$ by an appropriate tuning 
of the bare parameters. This can be done by fixing
 $\widehat\lambda$ and choosing $\widehat m$ to be in the symmetric region. 
As one 
decreases $\widehat m$, one gets closer to the critical line and the 
correlation
 length increases. Using this, one can reach the required correlation length.

To apply VSM we followed the same procedure as before. For six different 
$\widehat\lambda$'s and fixed correlation length $\zeta=3.6 \mp 4\%$,
 we calculated the values of $\bp_J$ for different values of $J$. The curve 
fitting 
procedure was carried out in the same way as for the previous cases. We 
noticed 
that in this regime the inclusion of larger $\phi_J$'s can change the behavior 
of the fit at small $\phi_J$, the region which is of most interest to us.
The problem arises due to the curve fitting procedure. In the 
WCR the data points close to $\ov \phi=0$ have much larger weighting that the 
one far away from this point. Thus, calculating the derivatives of $U(\bp)$
at $\ov \phi=0$ seem to be reliable. However, in the strong coupling regime, 
the data points that are far away from $\ov \phi=0$ have much higher weighting 
and even a small 
fluctuation might affect the calculated $J(\bp)$ considerably.

We improved the results by imposing the condition in Eq.~\eqn {cond}, that 
is to fixing the coefficient
 $a_1={1/\widehat G(0)}$ where $a_1$ is defined in Eq. \eqn {param}
and $\wt G(0)=N^2\mean{\wt \phi^2}$. This improved the results and the 
inclusion of larger
$\phi_J$'s did not affect the results significantly (up to $3\%$).

Next we calculated the renormalised parameters using CEPI. Unlike the 
previous cases the errors in the results were reasonable.
 For the extraction of renormalised parameters we only used two
histograms corresponding to $J=0$ and $J=0.005$. In the weak coupling regime 
where the mass term
is dominant, one needs to sample the higher values of $\bp$ in order to
improve the calculation of $\wh \lambda_r$. Thus in the strong coupling regime
there might not be a need for additional histograms. From the VSM results
 one might expect that sampling very high $\bp$ might have a similar problem.
 This was confirmed from our data for this particular case.

We also calculated the renormalised coupling using Eq.~\eqn {4vertex}. 
Unlike the  
weak coupling regime, uncertainties in the results in this region were 
reasonable. All the results in the 
strong coupling regime are shown in Fig. 5 and 6. They also are compared
with the strong coupling expansion results. They all seem to be in agreement 
with each other within errors. This indicates that as the coupling increases 
the MC results approach the strong coupling expansion results. In order to 
apply the CEPII method, there are 
numerous terms in (A.8) which have to be evaluated and consequently the 
accumulated 
errors can be very large. However we found that as before the renormalised 
mass can be calculated accurately. 

\subsection*{\bf Details of the simulations}

In our MC calculation we chose the Hybrid MC algorithm. In 
a run we have taken a number of decorrelation MC iterations between two 
measurements. All the calculations were done on a $20^2$ lattice 
and the rate of acceptance was kept between $40\%$ and $60\%$. In all cases
(except the broken sector) the calculations of renormalised mass and
 $\wt G(0)$ (where it was needed) and the direct calculation of 
$\wt\Gamma^{(4)}_c(0)$
were done using $6,800 $ uncorrelated samples with $50,000$ thermalisation 
configurations.  In the broken sector we used $11,000$ uncorrelated samples 
with 
the same thermalisation configurations. The reason for the increase was to 
obtain 
 better statistics, since the measured quantities have larger errors due to 
the 
non-vanishing disconnected pieces. 
In applying the VSM to the symmetric case (and in the weak coupling regime), 
we calculated $\bp$ with 
$0.025 \le J \le 0.425$.  We noticed that the necessary number of decorrelation
iterations in the presence of nonzero $J$ was smaller than for the $J=0$ case.
The calculations were carried out using $2,500$ decorrelated configurations. 
We took the number of thermalisation 
configurations to be $10,000$. In the broken symmetry sector we increased
the number of uncorrelated configurations to $3,200$.
In the strong coupling regime only the range of the values for $J$ was 
different (as mentioned in the previous section).
For CEPII we used $5,000$ uncorrelated configurations with $50,000$
thermalisation iterations. 
In construction of the probability distribution histograms, we used 
$750,000$ configurations. 
The curve fits were done using a standard $\chi^2$ fitting algorithm where
the uncertainties on the parameters were obtained from the diagonal of the 
covariant matrix.
 For the strong coupling results we also estimated the systematic 
error due to the fact
that $\widehat\zeta$ was fixed to be approximately $4\%$ by varying the fixed
value within reasonable limits.
\section {Conclusions}

We have studied the calculation of the effective potential for 
$\lambda\phi^4_{1+1}$ theory using three different methods: the variation 
of source method (VSM) and
two constraint effective potential methods (CEP I) and (CEP II). 
Using our new method, referred to as CEPII, we showed how to calculate the 
 vertex function using the correlation functions in the
 presence of a constraint. 
 We calculated the effective potential in the symmetric 
and the broken sector in the weak coupling regime as well as in the 
symmetric sector in the strong coupling regime. The renormalised quantities, 
$\wh\lambda_r$ and $\wh m_r$, were then obtained from the effective
potential for each case. In the weak coupling  regime we compared
 our results with  lattice perturbation theory. We found that in the 
symmetric case both VSM and CEP II can give accurate results, whereas the CEPI
method and the direct Monte Carlo calculation of the (2 and 4 point) vertex
functions failed to do so. We also found that in the
broken symmetry sector VSM is the 
most practical and accurate of these methods.
We also studied the model in the strong
 coupling regime and the results were compared with the strong coupling
 expansion results. In this regime we found 
 that CEPI, VSM, and the results from the
 direct Monte Carlo calculation of the vertex functions were
consistent with each other and with the
 strong coupling expansion results. In summary then, we have shown that
 Monte Carlo effective potential
 methods can be accurate
 and reliable tools for calculating physical quantities for  scalar field
 theories, but that one should use the method of evaluating the effective
potential and its derivatives which is best suited to the regime of interest.


\begin{appendix}
\section{}
The differential equations relating the constraint effective potential 
 and the classical potential are :\\

For the first derivative we have 
\be
{dU(\overline\phi) \over d\overline\phi }= \left\langle {dV(\phi) \over d\phi}
\right \rangle_{\overline\phi},
\ee \\
which for $\lambda\phi^4$ theory becomes
\be
{dU(\bp) \over d\bp}= m^2\bp +{\lambda \over 6}\mean{\phi^3}.
\ee
\newline
The second derivative is given by
\be
{d^2U(\overline\phi) \over d\overline\phi^2} = \left \langle {d^2V(\phi_1) 
\over d\phi_1^2}\right
 \rangle_{\overline\phi} - \sum_{i=1}^{N^d}  \left \langle  {dV(\phi_1) 
\over d\phi_1 } {dV(\phi_i) \over d\phi_i} \right \rangle_{\bp} +N^d 
\left(\left\langle
{dV(\phi_1) \over d\phi}\right\rangle_{\overline \phi}\right)^2,
\ee
which for $\lambda\phi^4$ theory becomes
\ba
{d^2U(\overline\phi) \over d^2\overline\phi} &=&
  m^2 +{\lambda \over 2}\mean{\phi^2}+N^d\left [
{\lambda^2 \over 36}\mean{\phi^3}^2+{\lambda m^2\over 3}\overline \phi \mean{
\phi^3}+m^4 \overline \phi ^2  \right] \nn
&&-N^d\left[m^4\mean{\phi\wt\phi}+{\lambda m^2  \over 6}\mean{\phi^3\wt\phi}+
{\lambda^2 \over 36}\mean{\phi^3\wt \phi^3} +{\lambda m^2  \over 6}
\mean{\phi \wt\phi^3} \right].
\label{app4}
\ea
The third derivative gives
\be
{d^3U(\bp) \over d\bp^3}= N^d {d^2U(\bp) \over d\bp^2}{dU(\bp) \over d\bp}
+2N^d \left ({dU(\bp) \over d\bp}\right )^2 
-N^{2d}\left({dU(\bp) \over d\bp}\right)^3 +
\left \langle {d^3V(\phi_1) \over d\phi^3_1}\right \rangle _{\bp}+\dots ,\\
\ee
\newline
which for $\lambda\phi^4$ theory becomes
\ba
{d^3U(\bp) \over d\bp^3}&=& \wh m^2 {\wh \lambda\over 6}\mean {\phi^3}-
N^d{\wh\lambda\over 6}
\mean {\phi^2{\wt \phi}^3}- N^d\wh \lambda 
\mean{\phi^2{\wt\phi}}  +\nn
 \left[{\cal O}({\wh\lambda}^2)\ {\rm terms}\right] &+&\left[{\cal O}(\wh m^2)
 \ {\rm terms }\right]+ \ 
 \left[{\rm terms \  that \ vanish \ at \  \bp=0}\right]+\dots
\ea
Finally for the fourth-derivative we obtain 
\be
{d^4U(\bp) \over d\bp^4}= N^d \left({d^2U(\bp)\over d\bp^2}\right)^2 -4N^{2d}
{d^2U(\bp)\over d\bp^2} \left({dU(\bp) \over d\bp}\right)^2    
+ \left \langle {d^4V(\phi_1) \over d\phi_1}\right \rangle _{\bp}+\dots ,\nn
\ee
which for $\lambda \phi^4$ theory gives
\ba
{d^4U(\bp) \over d\bp^4}&=&\wh \lambda-\lambda \wh m^2-N^d \lambda \wh m^2
\mean{{\wt \phi^2}}-N^d\wh \lambda \wh m^2\mean{\phi^2}\nn
+\left[{\cal O}(\lambda^2) \ {\rm terms }\right]&+&\left[{\cal O}(\wh m^4) 
\ {\rm terms}\right]+ \ 
 \left[{\rm terms \  that \ vanish \ at \  \bp=0}\right].
\label{app8}
\ea
\newline
\end {appendix}
\begin{table}
\caption{The comparison of the calculations of $\bp_J$ for different 
values of $J$'s and 
perturbative calculations of $\bp_J$ around $J=0.1$, $J=0.225$ and $J=0.425$
 with $\wh m^2=0.1$, $\wh \lambda=0.055$ and $N=20$.} 

\begin {center}
\begin{tabular}{lllll}
 $J$ & $\bp_J$ & error & $[\bp_J]_{\rm per}$ & error \\ \hline 
 0.050 & 0.4052 & 0.0032 & 0.409 & 0.0057 \\ 
 0.075 & 0.5841 & 0.0031 & 0.5880 & 0.0052 \\ 
 0.100 & 0.7648 & 0.0031 & 0.7648  & 0.0031\\ 
 0.125 & 0.9253 & 0.0027 & 0.9320 & 0.0042 \\ 
0.150 & 1.0840 &0.0027 & 1.095& 0.0048    \\                     
 0.175 & 1.2112 & 0.0027 & 1.218 & 0.0047      \\ 
 0.200 & 1.3399 & 0.0027 & 1.4510 & 0.0039 \\ 
 0.225 & 1.4505 & 0.0026 &1.4505   & 0.0026 \\ 
 0.250 & 1.5590 & 0.0026 & 1.5646& 0.0038\\ 
 0.275 & 1.6659 & 0.0026 & 1.6680 & 0.0043 \\ 
 0.300 & 1.7620 & 0.0026 & 1.7720 & 0.0049 \\ 
 0.325 & 1.8564 & 0.0026 & 1.8706 & 0.0048 \\ 
 0.350 & 1.9411 & 0.0026 & 1.945 & 0.0038 \\ 
 0.375 & 2.0166 & 0.0026 & 2.0209 & 0.0033 \\ 
 0.400 & 2.0963 & 0.0026 & 2.0963 & 0.0026 \\ 
 0.425 & 2.1704 & 0.0026 & 2.1714 & 0.0029 \\ 
 0.450 & 2.2430 & 0.0026 & 2.2470 & 0.0029 \\ 
\end{tabular}
\end {center}
\end{table}

\begin{table}
\caption{The comparison of the calculations of $\widehat\lambda_r$ using 
the VSM and the CEPII in the symmetric sector and weak coupling regime.
$ \wh \lambda_r^{cons}$ denotes the renormalised coupling calculated by 
CEPII. Here $\wh m^2=0.1$ and $N=20$.}
\begin {center}
\begin{tabular}{lllll}  
 $\widehat\lambda$ & $\widehat\lambda_r^{s}$ & error & 
$\widehat\lambda^{\rm cons}_r$ & error  \\ \hline
 0.02 & 0.0191 & 0.0003 & 0.018 & 0.0007 \\ 
 0.04 & 0.0386 & 0.0003 & 0.0363 & 0.0008 \\ 
 0.055 & 0.0518 & 0.0008 & 0.0510 & 0.0017 \\ 
 0.07 & 0.0670 & 0.0009 & 0.0657 & 0.0019 \\ 
0.1 & 0.0891 & 0.0008& 0.092& 0.0016 \\ 
 0.13 & 0.112 & 0.0013 & 0.121 & 0.004 \\ 
 0.19 & 0.165 & 0.002 & 0.175 & 0.0061 \\ 
 0.24 & 0.216 & 0.0023 & 0.22 & 0.007 \\ 
 0.35 & 0.313 & 0.0035 & 0.321 & 0.018 \\ 
\end{tabular}
\end {center}
\end{table}

\begin{table}
\caption{The comparison of the calculations of $\widehat m^2_r$ using 
the VSM and the CEPII in the symmetric sector and weak coupling regime. 
$\wh \lambda_r
^{cons}$ denotes the renormalised mass calculated by CEPII.
Here $\wh m^2=0.1$ and $N=20$.}

\begin {center}
\begin{tabular}{lllll} 
 $\widehat\lambda$ & $\widehat m_r$ & error & $\widehat m_r^{\rm cons}$ &
error \\ \hline
 0.02 & 0.324 & 0.001 & 0.323 & 0.007 \\ 
 0.04 & 0.334 & 0.002 & 0.330 & 0.007 \\ 
 0.055 & 0.340 & 0.0008 & 0.332 & 0.008\\ 
 0.07 & 0.343 & 0.0014 & 0.339 & 0.006 \\ 
0.1 & 0.345 &0.0023 & 0.347& 0.008    \\                     
 0.13 & 0.350 & 0.0023 & 0.357 & 0.009 \\ 
 0.19 & 0.375 & 0.0025 & 0.372 & 0.009 \\ 
 0.24 & 0.398 & 0.003 & 0.384 & 0.010 \\ 
 0.3 & 0.408 & 0.003 & 0.399 & 0.009 \\ 
 0.35 & 0.421 & 0.0035 & 0.410 & 0.010 \\ 
 0.40 & 0.433 & 0.004 & 0.428 & 0.010 \\ 
\end{tabular}
\end {center}
\end{table}

\begin{figure}[htb]
\epsfysize=8.5cm
\centering{\ \epsfbox{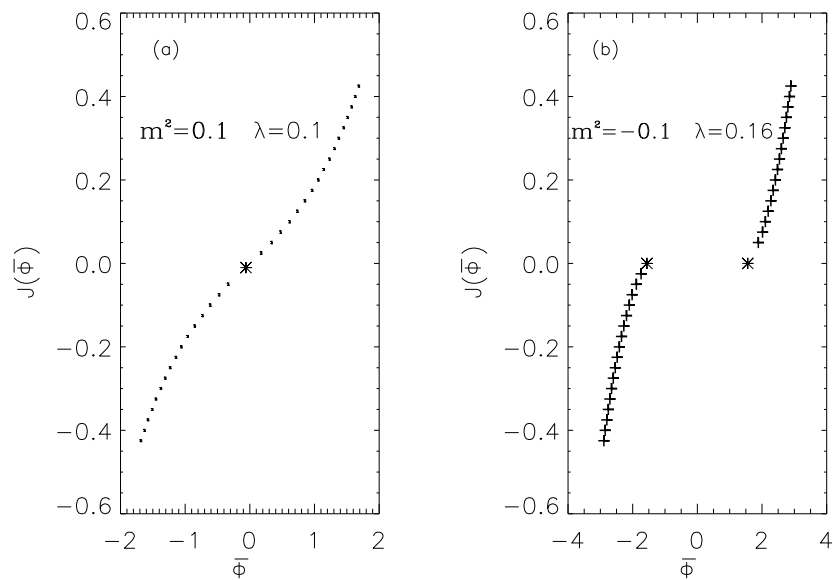}}
\caption{ An examples of $J(\bp)$ versus $\bp$ in the symmetric sector
 (a) and 
 for the broken symmetry sector (b). The stars (*) correspond to the values of
 $\bp$ at $J=0$. For these results $N=20$.}
\end{figure}

\begin{figure}[htb]
\epsfysize=6.5cm
\centering{\ \epsfbox{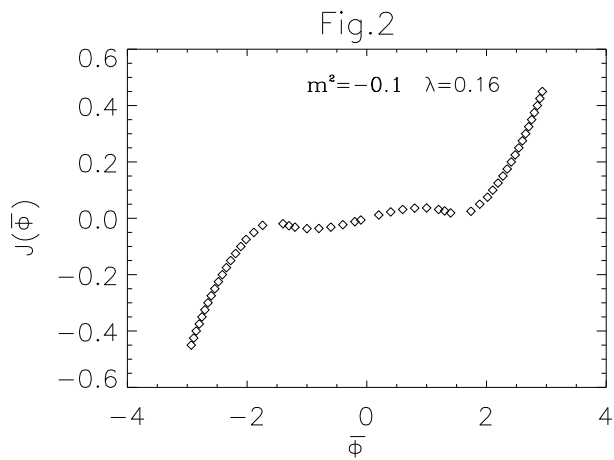}}
\caption{ An example of $J(\bp)$ versus $\bp$ in the broken sector using the 
constraint effective potential. For these results $N=20$.\newline\newline}
\end{figure}

\begin{figure}[htb]
\epsfysize=7.5cm
\centering{\ \epsfbox{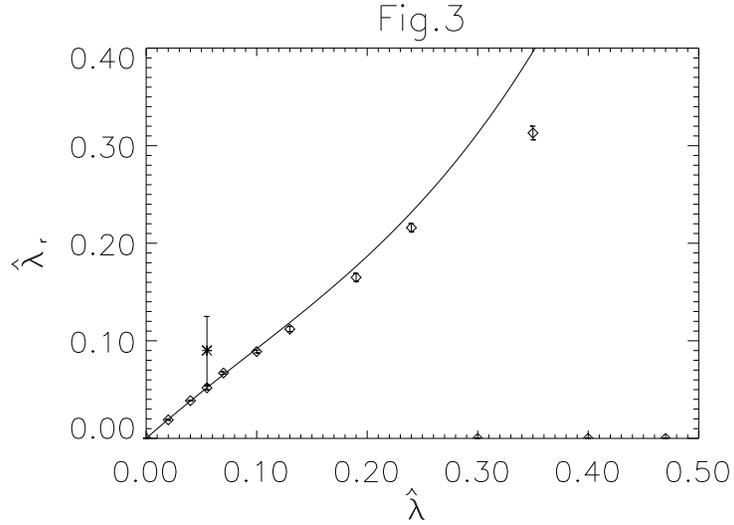}}
\caption{ The plot of $\widehat\lambda_r$ versus $\widehat\lambda$ in the
symmetric sector using lattice perturbation theory (solid line)
and  Eq. (40) (stars) and the VSM  (diamonds) with $\wh m^2=0.1$
and $N=20$.\newline}

\end{figure}

\begin{figure}[htb]
\epsfysize=7.5cm
\centering{\ \epsfbox{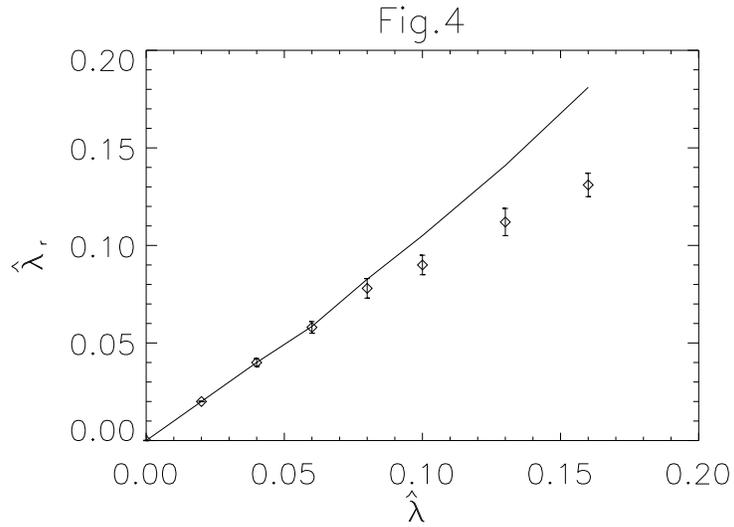}}
\caption{The plot of $\widehat\lambda_r$ versus $\widehat\lambda$ in the
broken symmetry sector
$\widehat\lambda$
 using lattice perturbation theory (solid line) and VSM (diamonds)
with $\wh m^2=-0.1$ and $N=20$.}
\end{figure}

\begin{figure}[htb]
\epsfysize=7.5cm
\centering{\ \epsfbox{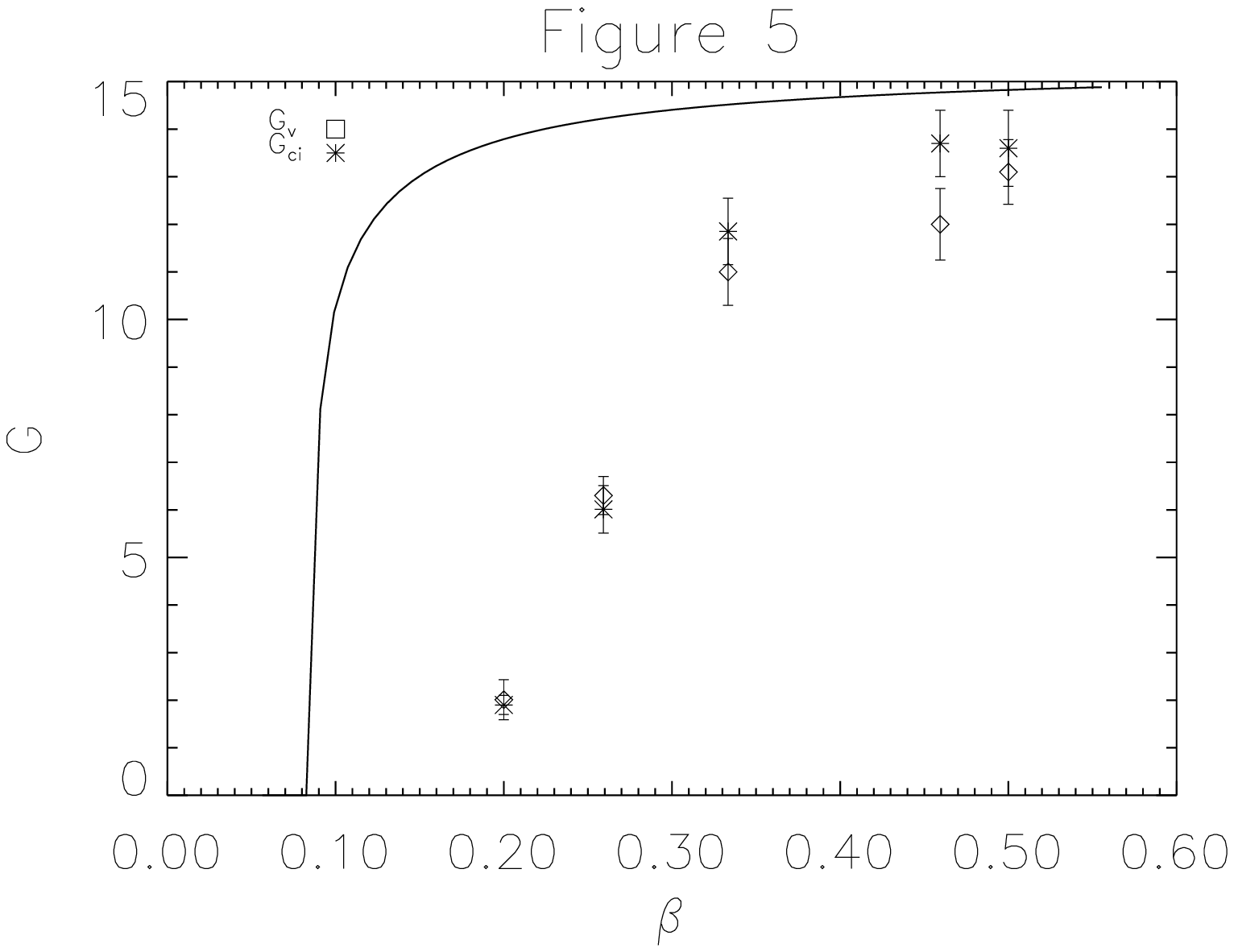}}
\caption{ The plot of $G={\widehat\lambda_r / \widehat m_r}$ versus 
$\beta= {\widehat\lambda+100 / \widehat\lambda}$ with strong coupling 
expansion results (solid line), using Eq. (40) (stars) and the VSM results 
(diamonds) with $\wh m^2_r=0.078 \pm 4\%$ and $N=20$.\newline\newline}
\end{figure}

\begin{figure}[htb]
\epsfysize=7.5cm
\centering{\ \epsfbox{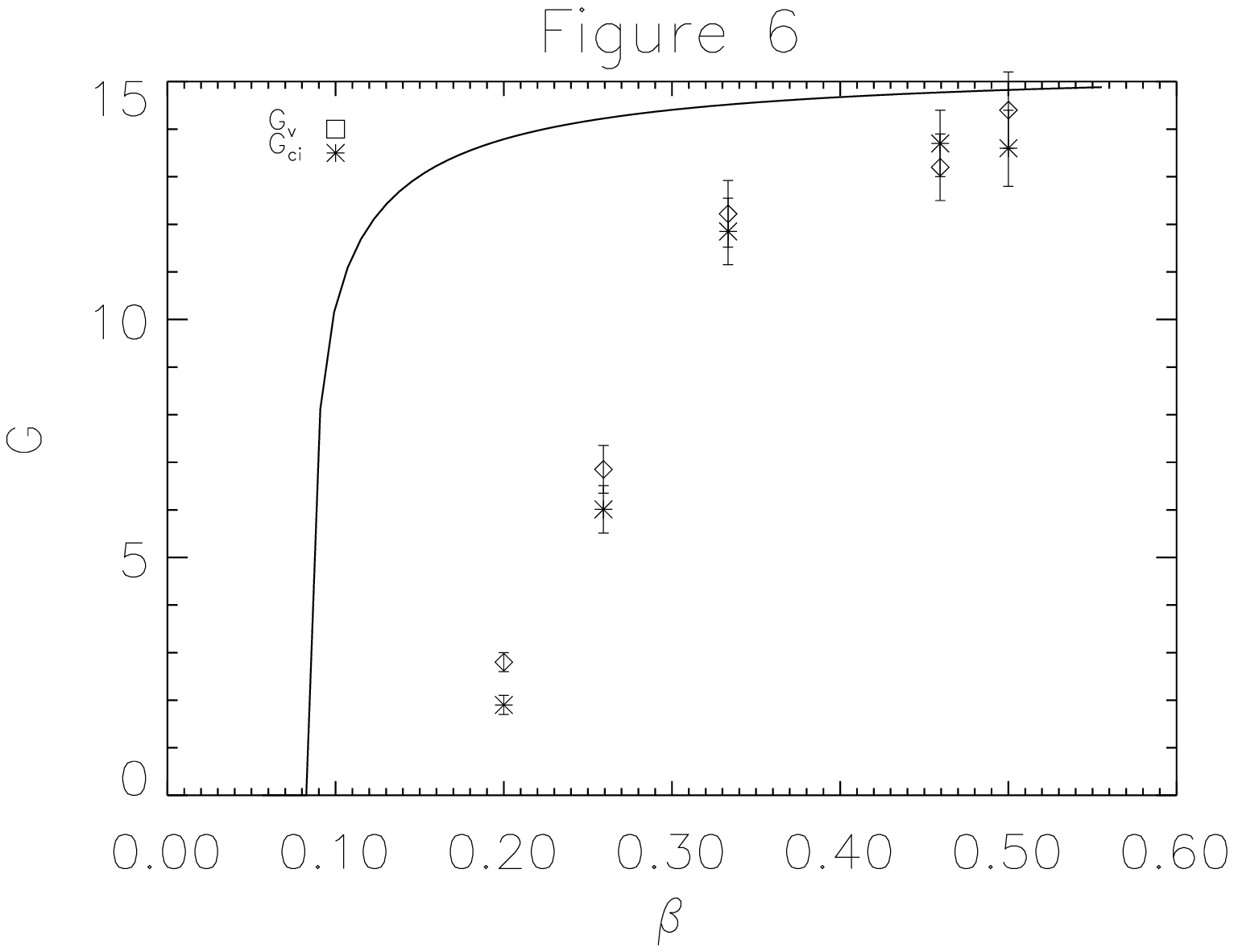}}
\caption{ The plot of $G={\widehat\lambda_r / \widehat m_r}$ versus 
$\beta= {(\widehat\lambda+100) / \widehat\lambda}$ with strong coupling 
expansion results (solid line), using 
Eq. (40) (stars) and the CPII method results (diamonds) with 
$\wh m^2_r=0.078 \pm 4\%$ and $N=20$.\newline\newline}
\end{figure}

\end{document}